\documentclass[a4paper,twocolumn,english,british,prb,showpacs]{revtex4}
\usepackage[latin9]{inputenc}
\usepackage{textcomp}
\usepackage{amstext}
\usepackage{amssymb}
\usepackage{graphicx}
\usepackage{esint}

\makeatletter

\pdfpageheight\paperheight
\pdfpagewidth\paperwidth

\@ifundefined{textcolor}{}
{%
 \definecolor{BLACK}{gray}{0}
 \definecolor{WHITE}{gray}{1}
 \definecolor{RED}{rgb}{1,0,0}
 \definecolor{GREEN}{rgb}{0,1,0}
 \definecolor{BLUE}{rgb}{0,0,1}
 \definecolor{CYAN}{cmyk}{1,0,0,0}
 \definecolor{MAGENTA}{cmyk}{0,1,0,0}
 \definecolor{YELLOW}{cmyk}{0,0,1,0}
}

\makeatother

\usepackage{babel}
\begin{document}

\title{Spectral edge mode in interacting one-dimensional systems}

\author{O. Tsyplyatyev}

\affiliation{School of Physics and Astronomy, The University of Birmingham, Birmingham,
B15 2TT, UK}

\author{A. J. Schofield}

\affiliation{School of Physics and Astronomy, The University of Birmingham, Birmingham,
B15 2TT, UK}

\date{\today}
\begin{abstract}
A continuum of excitations in interacting one-dimensional systems
is bounded from below by a spectral edge that marks the lowest possible
excitation energy for a given momentum. We analyse short-range interactions
between Fermi particles and between Bose particles (with and without
spin) using Bethe-Ansatz techniques and find that the dispersions
of the corresponding spectral edge modes are close to a parabola in
all cases. Based on this emergent phenomenon we propose an empirical
model of a free, non-relativistic particle with an effective mass
identified at low energies as the bare electron mass renormalised
by the dimensionless Luttinger parameter $K$ (or $K_{\sigma}$ for
particles with spin). 

The relevance of the Luttinger parameters beyond the low energy limit
provides a more robust method for extracting them experimentally using
a much wide range of data from the bottom of the one-dimensional band
to the Fermi energy. The empirical model of the spectral edge mode
complements the mobile impurity model to give a description of the
excitations in proximity of the edge at arbitrary momenta in terms
of only the low energy parameters and the bare electron mass. Within
such a framework, for example, exponents of the spectral function
are expressed explicitly in terms of only a few Luttinger parameters.
\end{abstract}


\maketitle

\section{Introduction}

The low energy properties of interacting particles in one-dimension
are well-described by the Tomonaga-Luttinger model\cite{TLModel}
based on the linear approximation to the spectrum of the excitation
at the Fermi energy. In this framework various correlation functions,
that involve a continuum of many-body excitations, can be evaluated
explicitly resulting in a common power-law behaviour - in contrast
to higher dimensions where the Fermi gas approximation with renormalised
parameters (the Fermi liquid model)\cite{Nozieres} remains robust.
In the last few decades different experimental realisations of one-dimensional
geometries were developed: carbon nanotubes,\cite{Nanotubes} cleaved
edge\cite{Yacoby_wires} or gated\cite{Ford_wires} one-dimensional
channels in semiconductor heterostructures, and cold atomic gases
in cigar shaped optical lattices\cite{ColdAtomWires} where the predictions
of the low energy theory\cite{Giamarchi} have already been observed
and measurements of high-energy effects are already possible.

Recently, a new theoretical understanding of the behaviour at high
energies was achieved by making a connection between the features
of the dynamical response of the one-dimensional systems and the Fermi
edge singularity in x-ray scattering in metals.\cite{Khodas} Application
of the mobile impurity model\cite{MobileImpurity} to the Tomonaga-Luttinger
model gives a description of excitations at high energies incorporating
dispersion of the spectral edge as an input parameter; the edge marks
the smallest excitation energy at a fixed momentum. Within the resulting
theory correlation functions exhibit a common power law behaviour
where exponents are related to the curvature of the spectral edge
and the Luttingers parameters.\cite{Glazman09,Pereira08,Schmidt10}
However, the theory for the edge mode itself remains an open problem. 

\begin{figure}
\begin{center}\includegraphics[width=0.9\columnwidth]{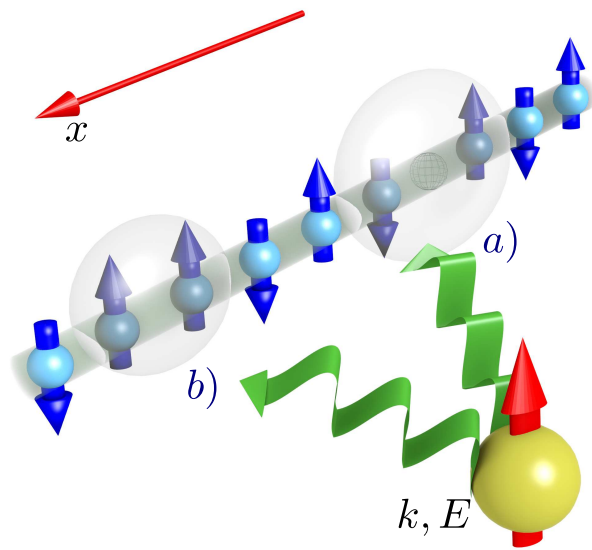}

\end{center}

\caption{A schematic representation of a tunnelling process into a one-dimensional
system for a particle with fixed momentum $k$ and energy $E$ that
is described by the spectral function. Excitations of the system are
a) density waves or a) and b) spin wave for particles with spin.}
\end{figure}
In this paper we analyse fundamental models of Fermi and Bose particles
with short-range interactions (with and without spin) in one dimension
using the available diagonalisation methods based on Bethe-Ansatz.
We investigate the edge mode of the spectral function - a dynamical
response function that generalises the single particle spectrum to
the many particle systems - and find that its dispersion is close
to a parabola for all cases in the thermodynamic limit.\cite{Repulsion}
It is exactly parabolic for fermions without spin and the biggest
deviation ($\lesssim20\%$) occurs for fermions with spin and a very
large interaction potential. Based on this result we propose an empirical
model of a free, non-relativistic particle for the spectral edge mode,
which describes a charge wave in the spinless case and a spin wave
in the spinful case, see a graphical representation of the spectral
function in Fig. 1. The effective mass $m^{*}$ is identified at low
energies as the bare electron mass $m$ strongly renormalised by the
dimensionless Luttinger parameter; $m^{*}/m=K$ and $m^{*}/m=K_{\sigma}$
in the spinless and the spinful case respectively. The position of
the edge of the spectral function in terms of this empirical model
can be expressed as 

\begin{equation}
\varepsilon_{\textrm{edge}}\left(k\right)=\mu+\frac{k_{F}^{2}}{2m^{*}}-\frac{\left(k-k_{0}\right)^{2}}{2m^{*}},\label{eq:edge_general}
\end{equation}
where $\mu$ is the chemical potential, $k_{F}$ is the Fermi momentum,
and $k_{0}=0\left(k_{F}\right)$ for Fermi (Bose) particles. 

The empirical model in Eq. (\ref{eq:edge_general}) breaks down when
the effective mass becomes infinite. At low energies $m^{*}=\infty$
is equivalent to zero sound velocity of the collective modes $v\left(v_{\sigma}\right)$.
The characteristic threshold is given by the quantum of the momentum
$v_{1}=2\pi/\left(mL\right)$ in a system of a finite size $L$. For
slower velocities $v\left(v_{\sigma}\right)\lesssim v_{1}$ the dispersion
of the spectral edge mode is not parabola-like and is not universal.

The parabolic shape of the spectral edge mode, which emerges in microscopic
calculation for different models, can be interpreted as an unusual
manifestation of Galilean invariance. The kinetic energy of a single
free particle is a parabolic function of its momentum, enforced by
the translational symmetry. Finite system size discretises the boosts
for changing inertial frames of reference in quanta of $2\pi/L$.
For a system consisting of $N$ particles the minimal boost of $2\pi N/L$
corresponds to the $2k_{F}$-periodicity in the momentum space; note
that interaction potentials are also Galilean invariant. However,
the total momentum of the whole many-particle system is still quantised
in the units of $2\pi/L$ that can be facilitated by giving a boost
to only a fraction of the particles $j<N$. The state on the spectral
edge with the momentum $k=2\pi j/L$ corresponds to a hole left between
$N-j$ particles in the rest frame and $j$ particles which have received
the minimal boost, see section III for details. The effective mass
of the hole-like quasiparticle is strongly renormalisation by interactions
since a partial boost is not a Galilean invariant transformation.
However, the parabolic dependence of the hole energy on momentum -
which is analogous to the kinetic energy of a free particle - is common
for different microscopic models thus it is an emergent phenomenon.

Excitations above the spectral edge are well-described at high energies
by the application of the mobile impurity model to the Tomonaga-Luttinger
theory which incorporates the curvature of the spectral edge as an
input parameter. \cite{GlazmanReview12} The result in Eq. (\ref{eq:edge_general})
removes this arbitrary input complementing the model above. Within
such a framework, for example, the edge exponents of the spectral
function are expressed explicitly in terms of only a few Luttinger
parameters and the bare electron mass that provides a systematic way
to classify them for a wide range of microscopic parameters.

The rest of the paper is organised as follows. Section II describes
the model of one-dimensional particles interacting via short range-potentials,
the corresponding spectral function, and discusses their general properties.
In section III we evaluate momentum dependence of the spectral edge
mode using the Bethe-Ansatz approach for Fermi particles in the fundamental
region. Section IV contains the effective field theory for excitations
above the spectral edge and calculates the edge exponents of the spectral
functions using the dispersion of the spectral edge mode itself obtain
in section III. In section V we show that Bose particles have the
same parabolic dispersion, with the mass renormalised by the same
Luttinger parameter $K$, of the spectral edge mode as the Fermi particles.
In section VI we summarise the results and discuss experimental implications.

\section{Model}

We consider particles in one-dimension interacting via a contact two-body
potential, $U$, as 

\begin{equation}
H=\int_{-\frac{L}{2}}^{\frac{L}{2}}dx\left(-\frac{1}{2m}\psi_{\alpha}^{\dagger}\left(x\right)\Delta\psi_{\alpha}\left(x\right)-UL\rho\left(x\right)^{2}\right)\label{eq:H}
\end{equation}
where $\psi_{\alpha}\left(x\right)$ are the field operators of Fermi
or Bose particles at point $x$ (with a spin $\alpha=\uparrow,\downarrow$
for spinful particles), $\rho\left(x\right)=\psi_{\alpha}^{\dagger}\left(x\right)\psi_{\alpha}\left(x\right)$
is the particle density operator, $L$ is the size of the system,
and $m$ is the bare mass of a single particle. Below we consider
periodic boundary conditions, $\psi_{\alpha}\left(x+L\right)=\psi_{\alpha}\left(x\right)$,
to maintain the translational symmetry of the finite length system,
restricting ourselves to repulsive interaction only, $U>0$, and we
assume $\hbar=1$.

The spectrum of excitations in the many-body case is given by the
spectral function which describes the response of a strongly correlated
system to a single particle excitation at energy $\varepsilon$ and
momentum $k$, $A_{\alpha}\left(k,\varepsilon\right)=-\textrm{Im}G_{\alpha\alpha}\left(k,\varepsilon\right)\textrm{sgn\ensuremath{\left(\varepsilon-\mu\right)}/\ensuremath{\pi}}$,
where $\mu$ is a chemical potential and $G_{\alpha\beta}\left(k,\varepsilon\right)=-i\int dxdte^{i\left(kx-\varepsilon t\right)}\left\langle T\left(e^{-iHt}\psi_{\alpha}\left(x\right)e^{iHt}\psi_{\beta}\left(0\right)\right)\right\rangle $
is a Fourier transform of Green function at zero temperature. To be
specific, we discuss particle like excitations, $\varepsilon>\mu$.
The spectral function in this domain reads \cite{AGD}

\begin{equation}
A_{\alpha}\left(k,\varepsilon\right)=\sum_{f}\left|\left\langle f|\psi_{\alpha}^{\dagger}\left(0\right)|0\right\rangle \right|^{2}\delta\left(\omega-E_{f}+E_{0}\right)\delta\left(k-P_{f}\right),\label{eq:A}
\end{equation}
where $E_{0}$ is the energy of the ground state $\left|0\right\rangle $,
and $P_{f}$ and $E_{f}$ are the momenta and the eigenenergies of
the eigenstate $\left|f\right\rangle $; all eigenstates are assumed
normalised. 

Galilean invariance defines a fundamental region for the spectrum
of excitations on the momentum axis. A minimal boost for changing
an inertial frame of reference for $N$ particles is $2\pi N/L$ which
is twice the Fermi momentum $k_{F}=\pi N/L$. In momentum space this
boost corresponds to $2k_{F}$-periodicity. We choose the fundamental
region as $-k_{F}<k<k_{F}$ for the Fermi and as $0<k<2k_{F}$ for
Bose particles. 

Under a $2k_{F}$-translation, the form factors in Eq. (\ref{eq:A})
do not change and the energies acquire simple shifts. The interaction
term in Eq. (\ref{eq:H}) is invariant under the transformation $x\rightarrow x+2\pi tj/\left(mL\right)$,
where $j$ is the number of the translation quanta, since the latter
can be absorbed into the a change of the integration variable. The
transformation of the momentum operator, $-i\nabla\rightarrow-i\nabla+2\pi j/L$,
in the the kinetic term results in a constant energy shift, $E\rightarrow E+2\pi jP/\left(mL\right)+2\pi^{2}j^{2}N/\left(mL^{2}\right)$,
of the Hamiltonian but keeps its matrix structure and, therefore,
eigenstates unaltered. Thus the spectral function can be extended
to arbitrary momenta by simultaneous translation of the momentum and
of the energy variables starting from the fundamental region.

Here, we are concerned with a distinctive feature of the spectral
function - the edge that marks the lowest possible excitation energy
for a given momentum. To identify its location we need to obtain only
the many-body spectrum of the model due to a singularity\cite{GlazmanReview12}
that guarantees large values of the form factors in the proximity
of the spectral edge. The two $\delta$-functions in Eq. (\ref{eq:A})
directly map the total momenta $P_{f}$ and the eigenenergies $E_{f}$
of all many-body states $\left|f\right\rangle $ into the points of
the spectral function $k$ and $\varepsilon$. We are going to identify
the states that have the smallest energy for each momentum and study
how the dispersion of the spectral edge mode, which they form, depends
on the interaction strength.

\section{Fermions}

\subsection{Spinless}

The zero range profile of two-body interaction potential in the model
in Eq. (\ref{eq:H}) has zero matrix elements for the Fermi particles
without spin due to the Pauli exclusion principle. A model of interactions
in this case requires a finite range of interactions which is usually
introduced by the point-splitting technique\cite{vonDelftReview}
developed to address the problem in the low energy limit. Here we
will use a different approach of introducing a lattice with the next-neighbour
interaction between particles. The lattice counterpart of the model
in Eq. (\ref{eq:H}) is the Hamiltonian $H=-\sum_{j=-L/2}^{L/2}\left(\psi_{j}^{\dagger}\psi_{j+1}+\psi_{j}^{\dagger}\psi_{j-1}\right)/\left(2m\right)-U\sum_{j=-L/2}^{L/2}\psi_{j}^{\dagger}\psi_{j}\psi_{j+1}^{\dagger}\psi_{j+1}$,
where $j$ is the site index on the lattice and the operators $\psi_{j}$
obey the Fermi commutation relations $\left\{ \psi_{i},\psi_{j}^{\dagger}\right\} =\delta_{ij}$. 

The model above can be diagonalised using the Bethe Ansatz approach.\cite{Korepin}
In the coordinate basis a superposition of $N$ plain waves, $\Psi=\sum_{P,j_{1}<\dots<j_{N}}e^{i\sum_{l=1}^{N}k_{P_{l}}j_{l}+i\sum_{l<l'=1}^{N}\varphi_{P_{l},P_{l'}}}\psi_{j_{1}}^{\dagger}\dots\psi_{j_{N}}^{\dagger}\left|\textrm{vac}\right\rangle $,
is an eigenstate, $H\Psi=E\Psi$, with the corresponding eigenenergy
\begin{equation}
E=\frac{1}{m}\sum_{j=1}^{N}\left(1-\cos\left(k_{j}\right)\right).\label{eq:E_fermions_nospin}
\end{equation}
Here $\left|\textrm{vac}\right\rangle $ is the vacuum state, the
scattering phases are fixed by the two-body scattering problem, 
\begin{equation}
e^{i2\varphi_{ll'}}=-\frac{e^{i\left(k_{l}+k_{l'}\right)}+1-2mUe^{ik_{l}}}{e^{i\left(k_{l}+k_{l'}\right)}+1-2mUe^{ik_{l'}}},\label{eq:2phi_fns}
\end{equation}
and $\sum_{P}$ is a sum over all permutations of $N$ quasimomenta.
The periodic boundary condition quantises the set of $N$ quasimomenta
simultaniously, 
\begin{equation}
Lk_{j}-2\sum_{l\neq j}\varphi_{jl}=2\pi I_{j},\label{eq:Ij_fns}
\end{equation}
where $I_{j}$ are a set of non-equal integer numbers. The total momentum
of $N$ particles, $P=\sum_{j}k_{j}$, is a conserved quantity.

\begin{figure}
\begin{center}\includegraphics[width=0.8\columnwidth]{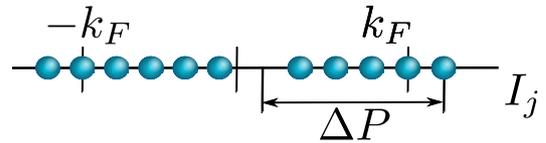}\end{center}\caption{\label{fig:spin-less-configuration}A set of quasimomenta from Eq.
(\ref{eq:quasimomenta_fermions_nospin}) that corresponds to the edge
mode of the spectral function for Fermi particles without spin. The
momentum of each many-particle state is given by $k=-k_{F}+\Delta P$.}
\end{figure}
The continuum model in Eq. (\ref{eq:H}) corresponds to the low density
(long wave length) limit of the lattice model. In this limit the scattering
phases in Eq. (\ref{eq:2phi_fns}) are linear functions of quasimomenta,
$2\varphi_{ll'}=\left(k_{l}-k_{l'}\right)/\left(1+\left(mU\right)^{-1}\right)+\pi$.
And the non-linear systems of equations in Eq. (\ref{eq:Ij_fns})
becomes linear. In the thermodynamic limit we solve it using perturbation
theory and obtain an independent quantisation condition for each quasimomentum
as solutions of the Bethe equations in the leading $1/N$-order,

\begin{equation}
k_{j}=\frac{2\pi I_{j}}{L-\frac{N}{1+\frac{1}{mU}}},\label{eq:quasimomenta_fermions_nospin}
\end{equation}
Thus all $N$-particles eigenstates can be labeled by all possible
sets of integers $I_{j}$ similarly to Slater determinants for free
fermions. The latter is possible as long as no bound states exist
- which is the case for any value of interaction strength $U\ge0$
in this limit.\cite{Korepin}

The eigenstates contributing to the spectral function satisfy the
number of particles constraint, \emph{i.e.} fixed to be $N+1$. The
lowest energy state for a fixed momentum $-k_{F}+\Delta P$ is given
by the set of integers in Fig. \ref{fig:spin-less-configuration}.
At low energies the system is in the universality class of Luttinger
liquids. Its properties are fully determined by the linear slope of
the spectrum of excitations at $\pm k_{F}$. Using the parameterisation
in Fig. \ref{fig:spin-less-configuration}, the first Luttinger parameter
(the sound velocity of the collective modes) is a discrete derivative
$v=L\left(E_{2}-E_{1}\right)/\left(2\pi\right)$, where $E_{2}$ and
$E_{1}$ are the energies of the states with $\Delta P=2\pi/L$ and
$\Delta P=0$. 

For Galilean invariant systems the product of the first and the second
(dimensionless $K$) Luttinger parameters gives the Fermi velocity
of the non-interacting system,\cite{Haldane81} $vK=v_{F}$ where
$v_{F}=\pi N/\left(mL\right)$. By a straightforward calculation of
the eigenenergies in Eq. (\ref{eq:E_fermions_nospin}) using Eq. (\ref{eq:quasimomenta_fermions_nospin})
for a pair states in Fig. \ref{fig:spin-less-configuration} with
$\Delta P=0,\;2\pi/L$ we directly obtain the second Luttinger parameter,
\begin{equation}
K=\left(1-\frac{N}{L\left(1+\frac{1}{mU}\right)}\right)^{2}.\label{eq:K_fermions_nospin}
\end{equation}

The dispersion of the spectral edge mode is given by the energies
and the momenta of all states in Fig. \ref{fig:spin-less-configuration}.
Starting from the solutions for quasimomenta in Eq. (\ref{eq:quasimomenta_fermions_nospin})
and repeating the same calculation as before, we directly obtain the
parabolic function of momentum\cite{OT13} in Eq. (\ref{eq:edge_general}),
where $m^{*}/m=K$ from Eq. (\ref{eq:K_fermions_nospin}). This calculation
also gives the chemical potential in Eq. (\ref{eq:edge_general})
as the bare electron mass renormalised by the Luttinger parameter
$K$, $\mu=k_{F}^{2}/\left(2mK\right)$.

\subsection{Spinful}

When Fermi particles have spin $1/2$, the Pauli exclusion principle
suppresses only the interaction between the particles with the same
spin orientation in the model in Eq. (\ref{eq:H}). The remaining
part of the density-density interaction term consists of a coupling
between particles with opposite spin orientations. 

This model can be diagonalised using the Bethe-Ansatz approach but
the Bethe hypothesis has to be applied twice.\cite{YG} In the coordinate
basis, a superposition of plain waves is an eigenstate, $H\Psi=E\Psi$,
of the model in Eq. (\ref{eq:H}),\begin{widetext} 
\begin{equation}
\Psi=\dotsint_{-\frac{L}{2}}^{\frac{L}{2}}dx_{1}\dots dx_{N}\sum_{P,Q}A^{PQ}e^{i\left(P\mathbf{k}\right)\cdot\left(Q\mathbf{x}\right)}\psi_{Q_{1}}^{\dagger}\left(x_{1}\right)\dots\psi_{Q_{N}}^{\dagger}\left(x_{N}\right)\left|\textrm{vac}\right\rangle ,\label{eq:Psi_fermions_spinfull}
\end{equation}
where the operators $\psi_{\alpha}\left(x\right)$ obey the Fermi
commutation rules $\left\{ \psi_{\alpha}\left(x\right),\psi_{\beta}^{\dagger}\left(x'\right)\right\} =\delta\left(x-x'\right)\delta_{\alpha\beta}$,
$k_{j}$ are $N$ quasimomenta, $\sum_{P,Q}$ is a sum over all permutations
of two independent sets of $N$ integer numbers ($P$ and $Q$), and
the coefficients $A^{PQ}$ are chosen by a secondary use of the Bethe
hypothesis, 
\begin{equation}
A^{PQ}=\textrm{sgn}\left(PQ\right)\sum_{R}\left(\prod_{1\leq l<l'\leq M}\frac{\lambda_{R_{l}}-\lambda_{R_{l'}}-imU}{\lambda_{R_{l}}-\lambda_{R_{l'}}}\right)\prod_{l=1}^{M}\frac{imU}{\lambda_{R_{l}}-k_{P_{z_{l}}}+\frac{imU}{2}}\prod_{j=1}^{z_{l}-1}\frac{\lambda_{R_{l}}-k_{P_{j}}-\frac{imU}{2}}{\lambda_{R_{l}}-k_{P_{j}}+\frac{imU}{2}}\;.
\end{equation}
\end{widetext}Here $\lambda_{l}$ are spin degrees of freedom of
$M$ ``up''-spins with respect to the reference ferromagnetic state
of $N$ ``down''-spins, $\sum_{R}$ is a sum over all permutations
of $M$ integer numbers, and $z_{l}$ is position of the $l^{\textrm{th}}$
spin $\uparrow$ in permutation $Q$. The eigenenergy corresponding
to the eigenstate in Eq. (\ref{eq:Psi_fermions_spinfull}) is 
\begin{equation}
E=\sum_{j=1}^{N}\frac{k_{j}^{2}}{2m}.\label{eq:E_fsf}
\end{equation}

The periodic boundary condition quantises the set of $N$ quasimomenta
$k_{j}$ (charge degrees of freedom) simultaneously, 
\begin{equation}
Lk_{j}-\sum_{l=1}^{M}\varphi_{jl}=2\pi I_{j},\label{eq:Bethe_Ij_fsf}
\end{equation}
where scattering phases $\varphi_{jl}=\allowbreak\log\Big[\left(\lambda_{l}-k_{j}-\frac{imU}{2}\right)\allowbreak/\left(\lambda_{l}-k_{j}+\frac{imU}{2}\right)\Big]\allowbreak/i$
depend on the quasimomenta of both kinds ($k_{j}$ and $\lambda_{l}$),
$I_{j}$ are a set of $N$ non-equal integer numbers, and $M$ quasimomenta
$\lambda_{l}$ (spin degrees of freedom) satisfy another set of non-linear
equations, 
\begin{equation}
\prod_{j=1}^{N}\frac{\lambda_{l}-k_{j}-\frac{imU}{2}}{\lambda_{l}-k_{j}+\frac{imU}{2}}=\prod_{l'=1\neq l}^{M}\frac{\lambda_{l'}-\lambda_{m}-imU}{\lambda_{l'}-\lambda_{m}+imU}.\label{eq:Bethe_Jl_fsf}
\end{equation}
The sum $P=\sum_{j=1}^{N}k_{j}$ is a conserved quantity - the total
momentum of $N$ particles.

The system of non-linear equations Eqs. (\ref{eq:Bethe_Ij_fsf}, \ref{eq:Bethe_Jl_fsf})
can be solved explicitly in the limit of infinite repulsion $U=\infty$.\cite{Ogata}
The quasimomenta $\lambda_{l}$ diverge in this limit. Under the substitution
of $\lambda_{l}=mU\tan y_{l}/2$, the second system of equations Eq.
(\ref{eq:Bethe_Jl_fsf}) becomes independent of the first system of
equations Eq. (\ref{eq:Bethe_Ij_fsf}) in leading $1/U$-order,

\begin{equation}
e^{iNy_{l}}=\left(-1\right)^{N+M-1}\prod_{l'=1\neq l}^{M}\frac{e^{i\left(y_{l}+y_{l'}\right)}+1+2e^{iy_{l}}}{e^{i\left(y_{l}+y_{l'}\right)}+1+2e^{iy_{l'}}}.\label{eq:Jl_Uinf_fsf}
\end{equation}
The above Bethe equations are identical to that of a Heisenberg anti-ferromagnet
\cite{Korepin} where the number of particles $N$ plays the role
of the system size. In one dimension a spin chain is mapped into the
model of interacting Fermi particles by the Jordan-Wigner transformation,
\cite{Korepin} Eq. (\ref{eq:Jl_Uinf_fsf}) are identical to Eqs.
(\ref{eq:2phi_fns}, \ref{eq:Ij_fns}) where the interaction strength
is set to $mU=-1$. Thus all solutions of Eq. (\ref{eq:Jl_Uinf_fsf})
can be labeled by all sets of $M$ non-equal integer numbers $J_{l}$
similarly to the case of Fermi particles without spin (see Fig. \ref{fig:spin-less-configuration}).
The system of equations for the quasimomenta $k_{j}$ in Eq. (\ref{eq:Bethe_Ij_fsf})
in the $U=\infty$ limit decouples into a set of single particles
quantisation conditions,

\begin{equation}
Lk_{j}=2\pi I_{j}+\frac{1-\left(-1\right)^{M}}{2}\pi+\sum_{l=1}^{M}y_{l}\label{eq:Ij_Uinf_fsf}
\end{equation}
Note that the independent magnetic subsystem, where quasimomenta $y_{l}$
satisfy Eq. (\ref{eq:Jl_Uinf_fsf}), is translationally invariant
thus $\sum_{l=1}^{M}y_{l}=2\pi\sum_{l=1}^{M}J_{l}/N$ (as can be checked
explicitly by multiplying Eq. (\ref{eq:Jl_Uinf_fsf}) for all $y_{l}$).
Therefore, the quantisation condition in Eq. (\ref{eq:Ij_1st_corr_fsf})
depends only on two sets of integer number $I_{j}$ and $J_{l}$.

\begin{figure}
\begin{center}\includegraphics[width=0.8\columnwidth]{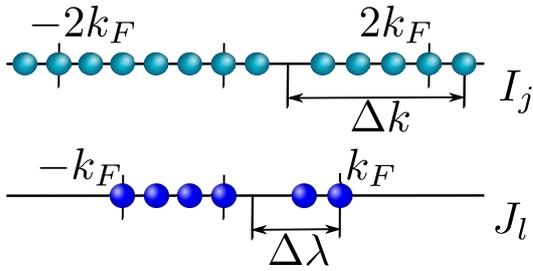}\end{center}

\caption{\label{fig:Spinfull-configuration}Parametrisation of many-body states
for fermions with spin using the $U=\infty$ limit in Eqs. (\ref{eq:Jl_Uinf_fsf},
\ref{eq:Ij_Uinf_fsf}). Charge-like excitations correspond to different
sets of $I_{j}$ and spin-like excitation correspond to different
sets of $J_{j}$.}
\end{figure}
All solutions of the original system of equations Eqs. (\ref{eq:Bethe_Ij_fsf},
\ref{eq:Bethe_Jl_fsf}) can be labeled by all sets of $N+M$ integer
numbers $I_{j}$ and $J_{l}$, see Fig. \ref{fig:Spinfull-configuration}.
The values of $k_{j}$ and $\lambda_{l}$ that correspond to these
integers can be obtained in two steps. Firstly, the spin degrees of
freedom $y_{l}$ that correspond to a set of $J_{l}$ are adiabatically
continued under a smooth deformation of Eq. (\ref{eq:Ij_fns}) from
$U=0$, which is the free particle limit, to $U=-1/m$, which coincides
with Eq. (\ref{eq:Jl_Uinf_fsf}). Note that the long wavelength solution
in Eq. (\ref{eq:quasimomenta_fermions_nospin}) can not be used here
because the most interesting case of zero polarisation for spinful
fermions corresponds to half-filling of the band for the model in
Eqs. (\ref{eq:Bethe_Ij_fsf}, \ref{eq:Bethe_Jl_fsf}) which is outside
of the limits of applicability of the low density regime. The values
$k_{j}$ that correspond to a set of $I_{j}$ and $J_{l}$ are obtained
directly from Eq. (\ref{eq:Ij_Uinf_fsf}). Secondly, the known values
of $k_{j}$ and $\lambda_{l}$ in the $U=\infty$ limit are adiabatically
continued under a smooth deformation of Eq. (\ref{eq:Bethe_Ij_fsf},
\ref{eq:Bethe_Jl_fsf}) to arbitrary value of the interaction strength
$U$. 

The interaction effects are controlled by a single dimensionless parameter
that can be defined using the $1/U$ corrections in the large $U$
limit. Power series expansion of the Eqs. (\ref{eq:Bethe_Ij_fsf},
\ref{eq:Bethe_Jl_fsf}) up to the first subleading $1/U$-order, $\lambda_{l}=mU\tan y_{l}/2+y_{l}^{\left(1\right)}$
and $k_{j}=k_{j}^{\left(0\right)}+2k_{j}^{\left(1\right)}/\left(mU\right)$,
where $y_{l}$ and $k_{j}^{\left(0\right)}$ are the solutions of
Eqs. (\ref{eq:Jl_Uinf_fsf}, \ref{eq:Ij_Uinf_fsf}) , yields
\begin{equation}
\sum_{j=1}^{N}\left(k_{j}^{\left(0\right)}-y_{l}^{\left(1\right)}\right)\cos^{2}y_{l}=-2\sum_{l'=1\neq l}^{M}\frac{y_{l'}^{\left(1\right)}-y_{l}^{\left(1\right)}}{\left(\tan y_{l}-\tan y_{l'}\right)^{2}+4},\label{eq:Jl_1st_corr_fsf}
\end{equation}
and
\begin{equation}
k_{j}^{\left(1\right)}=\frac{2}{L}\sum_{l=1}^{M}\left(k_{j}^{\left(0\right)}-y_{l}^{\left(1\right)}\right)\cos^{2}y_{l}.\label{eq:Ij_1st_corr_fsf}
\end{equation}
The first order coefficients $y_{l}^{\left(1\right)}$ can be expressed
from Eq. (\ref{eq:Jl_1st_corr_fsf}) in terms of zeroth order coefficients
$k_{j}^{\left(0\right)}$ and $y_{l}$. Then, in the thermodynamic
limit, the first order corrections to the quasimomenta $k_{j}$ in
Eq. (\ref{eq:Ij_1st_corr_fsf}) become $k_{j}^{\left(1\right)}=2k_{j}^{\left(0\right)}\sum_{m=1}^{M}\cos^{2}y_{l}/L$.
This gives a condition of validity for the $1/U$-expansion of the
Bethe equations, $2k_{j}^{\left(1\right)}/\left(mUk_{j}^{\left(0\right)}\right)$,
which is independent of both indices $j$ and $l$.

\begin{figure}[t]
\begin{center}\includegraphics[width=1\columnwidth]{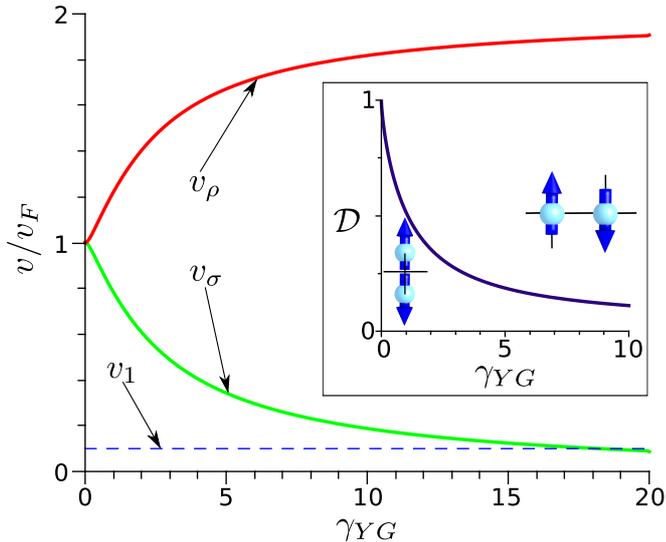}

\end{center}

\caption{\label{fig:low_energy_fsf}Velocities of the collective modes for
fermions with spins at low energies as a function of the interaction
parameters $\gamma_{YG}$ from Eq. (\ref{eq:gammaYG}). The red line
corresponds to the holon branch, the green line corresponds to the
spinon branch, and the blue dashed line marks a quantum of momentum
$v_{1}=2\pi/(mL)$; $L=400$, $N=40$, $\gamma_{YG}=3.44mU$. Inset:
Degree of double degeneracy, see the definition in Eq. (\ref{eq:Degeneracy}),
for the ground state on Fig. \ref{fig:Spinfull-configuration} with
$\Delta k=\Delta\lambda=0$ as a function of the interaction parameter
$\gamma_{YG}$.}
\end{figure}

We use the latter to define a single parameter, 
\begin{equation}
\gamma_{YG}=\frac{mL}{2N}\frac{U}{\left(1+\frac{1}{N}\sum_{l=1}^{M}\cos y_{l}\right)},\label{eq:gammaYG}
\end{equation}
that characterises the degree of repulsion between fermions. When
$\gamma_{YG}\gg1$ the particles with opposite spin orientations scatter
strongly off each other and when $\gamma_{YG}\ll1$ they interact
weakly with each other. For example, this is manifested in a change
of degeneracy of the quasimomenta $k_{j}$ that correspond to the
ground state of unpolarised Fermi particles, $M=N/2$. We account
for the degree of double degeneracy with respect to spin-1/2 using

\begin{equation}
\mathcal{D}=2-\frac{L\sum_{j=1}^{N-1}\left(k_{j+1}-k_{j}\right)}{\pi N}.\label{eq:Degeneracy}
\end{equation}
This quantity is $\mathcal{D}=1$ when each momentum state of free
fermions is doubly occupied ($U=0$) and is $\mathcal{D}=0$ when
each momentum state is occupied by a single particle ($U=\infty$).
The crossover from one regime to another occurs at $\gamma_{YG}=1$
where $\mathcal{D}$ crosses the value of $1/2$, see inset in Fig.
\ref{fig:low_energy_fsf}.

The ground state of the model in Eq. (\ref{eq:H}) has zero spin polarisation
when the external magnetic field is absent, $M=N/2$. To be specific
we consider the ground states with even values of $N$ and $M$. Excited
states contributing to the spectral function satisfy the number of
particles being constrained to be $N+1$. In the $U=\infty$ limit
the lowest energy eigenstates for a fixed momentum $P=-k_{F}+\Delta P$
are given by a set of integers in Fig. \ref{fig:Spinfull-configuration}
with $\Delta k=0$ and $\Delta P=\Delta\lambda$.\cite{Essler10}
In the opposite limit of free fermions, the lowest energy eigenstates
for a fixed momentum $P=-k_{F}+\Delta P$ are doubly degenerate with
respect to spin-1/2 and are given by the set of integers in Fig. \ref{fig:Spinfull-configuration}
for each spin orientation. The quasimomenta in both limits are smoothly
connected under adiabatic deformation of Eqs. (\ref{eq:Bethe_Ij_fsf},
\ref{eq:Bethe_Jl_fsf}) from $U=\infty$ to $U=0$ marking the edge
of the spectral function in Eq. (\ref{eq:A}) for arbitrary $U$. 

At low energies the eigenstate are strongly mixed in the spin sector
due to spin-charge separation\cite{Giamarchi} implying that $A_{\uparrow}\left(k,\varepsilon\right)=A_{\downarrow}\left(k,\varepsilon\right)$.
The excitations of the system are spinons and holons which are well
approximated by the spinful generalisation of the Tomonaga-Luttinger
model with only four free parameters $v_{\rho,\sigma}$ and $K_{\rho,\sigma}$.
The pair of velocities are the slopes of the linearised dispersions
of the charge and spin excitations at $\pm k_{F}$. Using the representation
of the eigenstates in Fig. \ref{fig:Spinfull-configuration} they
are

\begin{equation}
v_{\rho}=\frac{L\left(E_{2}-E_{1}\right)}{2\pi},\quad v_{\sigma}=\frac{L\left(E_{3}-E_{1}\right)}{2\pi},\label{eq:v_fsf}
\end{equation}
where $E_{1}$, $E_{2}$, and $E_{3}$ correspond to the energies
of the states with ($\Delta k=0$, $\Delta\lambda=0$), ($\Delta k=2\pi/L$,
$\Delta\lambda=0$), and ($\Delta k=0,\Delta\lambda=2\pi/L$) respectively.\cite{Coll}
The numerical evaluation of $v_{\rho,\sigma}$ as a function of the
interaction parameter $\gamma_{YG}$\cite{gamma_yg_gs} is presented
in Fig. \ref{fig:low_energy_fsf}. For $\gamma_{YG}=0$ both velocities
coincide $v_{\rho}=v_{s}=v_{F}$. For large $\gamma_{YG}\gg1$ the
holon velocity doubles $v_{\rho}=2v_{F}$ due to strong repulsion
between particles with opposite spin orientations\cite{Schultz} and
the spinon velocity becomes zero $v_{\sigma}=0$ since it vanishes
as $\sim1/\left(m^{2}U\right)$ in this limit.\cite{Korepin} The
other pair of Luttinger parameters can be obtained directly for Galilean
invariant systems using $v_{\rho,\sigma}$ and the Fermi velocity,
$K_{\rho,\sigma}=v_{F}/v_{\rho,\sigma}$ where $v_{F}=\pi M/L$, without
the need of a second observable such as compressibility.\cite{Giamarchi}

\begin{figure}[t]
\begin{center}\includegraphics[width=1\columnwidth]{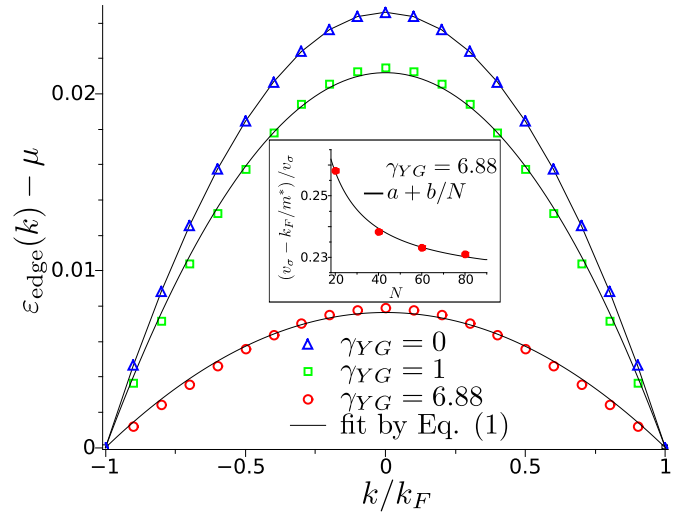}

\end{center}

\caption{\label{fig:Edge-for-unpolarised}Dispersion of the spectral edge mode
(extension of the spinon branch to high energies) for fermions with
spins for different values of the interaction parameter $\gamma_{YG}=0,1,6.88$;
$L=400$, $N=40$. The blue triangles, green squares, and red ellipses
are the numerical solutions of Eqs. (\ref{eq:Bethe_Ij_fsf}, \ref{eq:Bethe_Jl_fsf}),
the solid black lines are the best parabolic fits by Eq. (\ref{eq:edge_general}).
Inset: Difference between the slope of the parabolic dispersion at
$E_{F}$, which is given by the effective mass $m^{*}$, and the velocity
of spin waves at low energies, which is obtained directly from Eq.
(\ref{eq:v_fsf}) - $\left(v_{\sigma}-k_{F}/m^{*}\right)/v_{\sigma}$
- as a function of the number of particles $N$ for $\gamma_{YG}=6.88$.
The solid black line is the $1/N$ fit, \foreignlanguage{english}{$a+b/N$,
that gives }$a=0.22$.\foreignlanguage{english}{\cite{parabolic_fit_fsf}}}
\end{figure}
 Beyond the linear regime the position of the edge of the spectral
function is given by following of the low energy spinon mode. Numerical
evaluation shows that $\varepsilon_{\textrm{edge}}\left(k\right)=E_{k}-E_{0}$,
where $E_{k}$ corresponds to the states in Fig. \ref{fig:Spinfull-configuration}
with $\Delta k=0$ and $k=-k_{F}+\Delta\lambda$, is close to a parabola
for all values of $\gamma_{YG}$, see Fig. \ref{fig:Edge-for-unpolarised}.
For $\gamma_{YG}=0$ the shape of the spectral edge mode is exactly
parabolic following the dispersion of free Fermi particles. For $\gamma_{H}\gg1$
deviations from a parabola are largest. We quantify them by comparing
the effective mass $m^{*}$, obtained by the best fit of Eq. (\ref{eq:edge_general})
at all energies, with the spinon velocity $v_{\sigma}$ from Eq. (\ref{eq:v_fsf}),
obtained at low energy. The deviation \foreignlanguage{english}{$\left(v_{\sigma}-k_{F}/m^{*}\right)/v_{\sigma}$
decreases as the number of particles $N$ grows but it saturates at
a finite value of $\sim0.2$ in the limit $N\rightarrow\infty$,\cite{parabolic_fit_fsf}
see inset in Fig. }\ref{fig:Edge-for-unpolarised}\foreignlanguage{english}{. }

The edge of the spectral function in the complementary part of the
fundamental range, $k_{F}<k<3k_{F}$, also has a parabolic shape.
The eigenstates with the smallest eigenenergies for a fixed momentum
$k$ in this range are connected with their counterparts in the $-k_{F}<k<k_{F}$
range by a shift of the spin variables $\lambda_{j}\rightarrow\lambda_{j}+2\pi/L$.
Repeating the same numerical procedure as before for $\varepsilon_{\textrm{edge}}\left(k\right)=E_{k}-E_{0}$,
where $E_{k}$ corresponds to the states in Fig. \ref{fig:Spinfull-configuration}
with $\Delta k=0$ and $k=k_{F}+\Delta\lambda$, we obtain the result
in Eq. (\ref{eq:edge_general}) with $k_{0}=2k_{F}$. In the ``hole
region'' $\varepsilon<\mu$, the position of the edge of the spectral
function is obtained by reflection of $\varepsilon_{\textrm{edge}}\left(k\right)$
with respect to the line $\varepsilon=\mu$.

The parabola-like behaviour of the edge mode breaks down in finite
sized systems in the ultra-strong interaction regime when the spinon
velocity $v_{\sigma}$ becomes smaller than its own quantum set by
the finite size of the system \foreignlanguage{english}{$v_{1}=2 \pi/\left(mL\right)$,
see the dashed line in Fig. }\ref{fig:low_energy_fsf}. Correspondingly,
the threshold for entering this regime becomes $\gamma_{YG}\rightarrow\infty$
in the thermodynamic limit, as observed in Fig. \ref{fig:low_energy_fsf}
when $v_{1}\rightarrow0$. When \foreignlanguage{english}{$v_{\sigma}<v_{1}$},
the behaviour of the system is dominated by doubling of the period
in the momentum space from $2k_{F}$ to \foreignlanguage{english}{$4k_{F}$
which can be seen explicitly from Eqs. (\ref{eq:E_fsf}, \ref{eq:Ij_Uinf_fsf})
in the $U=\infty$ limit. The doubling in the spinful case is a direct
consequence of Galilean invariance of the model in Eq. (\ref{eq:H}).
However, it does not manifest itself in the thermodynamic limit for
finite spinon velocities $v_{\sigma}>v_{1}$, for which the edge of
the spectral function is still $2k_{F}$-periodic.}

\selectlanguage{english}%

\section{Effective field theory}

Eigenmodes above the spectral edge can be described by ``the mobile
impurity model''\cite{GlazmanReview12} with two different types
of fields that account for all possible low energy excitations with
respect to a state on the spectral edge with a given momentum $k$
in Figs. \ref{fig:spin-less-configuration} and \ref{fig:Spinfull-configuration}.
One field is responsible for bosonic excitations around $\pm k_{F}$
whose behavior is well-approximated by the Tomonaga-Luttinger model.
Another field models the dynamics of the hole-like degree of freedom,
as observed in Fig. \ref{fig:spin-less-configuration} for a large
$\Delta P$. For a $k$ away from $\pm k_{F}$ creation of a second
or removal the existing hole-like excitation is associated with a
significant energy cost thus the corresponding field describes a single
Fermi particle.

\selectlanguage{british}%
The interaction between the deep hole and the excitations at $\pm k_{F}$
is of the density-density type since their corresponding energy bands
are separated by a large barrier. Bosonisation of the excitations
at $\pm k_{F}$ leaves two unknown coupling constants between a pair
of the canonically conjugated variables of the Tomonaga-Luttinger
model and a fermionic field of the deep hole that can be identified
by considering two different physical properties.\cite{Glazman09,Kamenev}
One is translation invariance of the hybrid system that can be represented
as a motion of a fermionic excitation in a bosonic fluid with the
velocity $u=\left\langle \nabla\theta\right\rangle /m$. Another is
an observable that corresponds to the change of the total energy with
respect long-range variations of the density, which for the hybrid
systems is given by $\delta\rho=-\left\langle \nabla\varphi\right\rangle /\pi$.
Here $\varphi$ and \foreignlanguage{english}{$\nabla\theta$ are
the canonically conjugated variables of the Tomonaga-Luttinger model
that correspond to the density and the current of the hydrodynamic
modes respectively.}

\selectlanguage{english}%
For a fixed value of $k$, the dynamics of the free Bose-like and
the free Fermi-like fields can be linearised \foreignlanguage{british}{for
states close to the spectral edge. Using the dispersion in Eq. (\ref{eq:edge_general})
for the Fermi-like field and the Luttinger parameters for the Bose-like
field, the mobile impurity model reads\begin{widetext}
\begin{equation}
H=\int dx\left[\frac{v}{2\pi}\left(K\left(\nabla\theta\right)^{2}+\frac{\left(\nabla\varphi\right)^{2}}{K}\right)+\left(\frac{k\left(K-1\right)}{m^{*}}\nabla\theta+\frac{v\left(K+1\right)}{K}\nabla\varphi\right)d^{\dagger}d+d^{\dagger}\left(\frac{k^{2}}{2m^{*}}-\frac{i\nabla}{m^{*}}\right)d\right]\label{eq:H_eff_fns}
\end{equation}
\end{widetext}where $-k_{F}<k<k_{F}$ is the total momentum of the
system - an input parameter of the model, $m^{*}=mK$ is the effective
mass of the deep hole, $v$ and }$K$\foreignlanguage{british}{ are
the Luttinger parameters defined at $\pm k_{F}$, the fields $\theta$
and }$\varphi$\foreignlanguage{british}{ are the canonically conjugated
variables $\left[\varphi\left(x\right),\nabla\theta\left(y\right)\right]=i\pi\delta\left(x-y\right)$
of the Tomonaga-Luttinger model, and the field $d$ obey the Fermi
commutation rules $\left\{ d\left(x\right),d^{\dagger}\left(y\right)\right\} =\delta\left(x-y\right)$.}

\selectlanguage{british}%
The Hamiltonian in Eq. (\ref{eq:H_eff_fns}) can be diagonalised by
a unitary transformation.\cite{Khodas,Glazman09} The rotation $e^{-iU}He^{iU}$,
where $U=\int dy\left[C_{\pm}\left(\sqrt{K}\theta+\varphi/\sqrt{K}\right)\allowbreak+C_{\pm}\left(\sqrt{K}\theta-\varphi/\sqrt{K}\right)\right]d^{\dagger}d$
and $C_{\pm}=\allowbreak\left(2\sqrt{K}\right)^{-1}\allowbreak\Big[k\left(K-1\right)\allowbreak\pm k_{F}\left(K+1\right)\Big]\allowbreak/\left(k\pm k_{F}\right)$,
eliminates the coupling term between the fields turning Eq. (\ref{eq:H_eff_fns})
into a pair of free harmonic models. Then, the observables can be
calculated in a straightforward way as averages over free fields only. 

The spectral function in Eq. (\ref{eq:A}) can calculated using the
effective field model.\cite{Khodas} The original operators $\psi^{\dagger}\left(x\right)$
of Fermi particles of the model in Eq. (\ref{eq:H}) correspond to
a composite excitation consisting of two bosons and one fermion in
the field language of the model in Eq. (\ref{eq:H_eff_fns}) \foreignlanguage{english}{\emph{(}see
the state in Fig. \ref{fig:spin-less-configuration}}). The fermionic
excitation gives a dominant contribution to the spectral weight $\left|\left\langle f|\psi^{\dagger}\left(0\right)|0\right\rangle \right|^{2}$,
thus at leading order in \foreignlanguage{english}{$\left|\varepsilon-\varepsilon_{\textrm{edge}}\left(k\right)\right|$}
close to the spectral edge the spectral function reads $A\left(k,\varepsilon\right)=\int dtdxe^{i\left(kx-\varepsilon t\right)}\left\langle d^{\dagger}\left(x,t\right)d\left(0,0\right)\right\rangle $
where $d\left(x,t\right)=e^{-iHt}d\left(x\right)e^{iHt}$ and $\left\langle \dots\right\rangle $
is the zero temperature expectation value with respect to the model
in Eq. (\ref{eq:H_eff_fns}). In the diagonal basis the average is
evaluated over free fields by standard means. Following the steps
of Ref. \onlinecite{Glazman09} we obtain $A\left(\varepsilon,k\right)\allowbreak\sim\theta\left(\varepsilon-\varepsilon_{\textrm{edge}}\left(k\right)\right)\allowbreak/\left|\varepsilon-\varepsilon_{\textrm{edge}}\left(k\right)\right|^{-\alpha}$
where the exponent depends only on the Luttinger parameter $K$, 
\begin{equation}
\alpha=1-\frac{K}{2}\left(1-\frac{1}{K}\right)^{2}.\label{eq:alpha_fns}
\end{equation}
This result is the same for the particle and the hole parts of the
spectrum. Here $K$ is given by the analytic result in Eq. (\ref{eq:K_fermions_nospin}).

Excitations above the spectral edge for Fermi particles with spin
can be described using the mobile impurity model in an analogous way.\cite{Schmidt10,Essler10}
The number of the bosonic fields doubles due to the two spin orientations.
Bosonisation of the modes at $\pm k_{F}$ gives a diagonal Tomonaga-Luttinger
model in the basis of spin and charge fields. Here there are four
unknown coupling constants between two pair of the canonically conjugated
variables of the Tomonaga-Luttinger model and the Fermi-like field
of the deep hole. One pair of the constants that corresponds to the
coupling to spinon modes are zero due to the symmetry with respect
to the spin orientation in the original microscopic model in Eq. (\ref{eq:H}),
where the external magnetic field is zero. Another pair of the constants
that correspond to the coupling to holon modes can be identified by
considering the same physical properties as for the Fermi particles
without spin. 

Using the result in Eq. (\ref{eq:edge_general}) and the Luttinger
parameters, the mobile impurity model reads\begin{widetext}
\begin{equation}
H=\int dx\Bigg[\sum_{\alpha=\rho,\sigma}\frac{v_{\alpha}}{2\pi}\left(K_{\alpha}\left(\nabla\theta_{\alpha}\right)^{2}+\frac{\left(\nabla\varphi_{\alpha}\right)^{2}}{K_{\alpha}}\right)+\frac{v_{\sigma}-\frac{k}{m^{*}}}{\sqrt{2}}\left(K_{\sigma}\nabla\theta_{\rho}+\nabla\varphi_{\rho}\right)d^{\dagger}d+d^{\dagger}\left(\frac{k^{2}}{2m^{*}}-\frac{i\nabla}{m^{*}}\right)d\Bigg],\label{eq:H_eff_fsf}
\end{equation}
\end{widetext}where $k$ is the total momentum of the system - an
input parameter of the model, $m^{*}=mK_{\sigma}$ is the effective
mass of the deep hole, $v_{\rho}$, $K_{\rho}$, $v_{\sigma}$, $K_{\sigma}$
are the four Luttinger parameters for the spin and the charge modes,
the bosonic fields $\theta_{\rho}$,$\varphi_{\rho}$,$\theta_{\sigma}$,$\varphi_{\sigma}$
are canonically conjugated variables $\left[\varphi_{\alpha}\left(x\right),\nabla\theta_{\beta}\left(y\right)\right]=i\pi\delta_{\alpha\beta}\delta\left(x-y\right)$
of the Tomonaga-Luttinger model, and the field $d$ obey the Fermi
commutation rules $\left\{ d\left(x\right),d^{\dagger}\left(y\right)\right\} =\delta\left(x-y\right)$.

The diagonalisation of the Hamiltonian in Eq. (\ref{eq:H_eff_fsf})
can be done by a unitary transformation in a very similar fashion
to the spinless case.\cite{Schmidt10,Essler10} The rotation $e^{-iU}He^{iU}$,
where $U=\int dx\Big[C_{+}\Big(\sqrt{K_{\rho}}\theta+\varphi/\sqrt{K_{\rho}}\Big)\allowbreak+C_{-}\left(\sqrt{K_{\rho}}\theta-\varphi/\sqrt{K_{\rho}}\right)\Big]\allowbreak d^{\dagger}d$
and $C_{\pm}=\mp\sqrt{K_{\rho}}8^{-\frac{5}{2}}\allowbreak\left(k-k_{F}\right)\allowbreak\left(K_{\rho}^{-1}\mp K_{\sigma}^{-1}\right)\allowbreak/\left(k/K_{\sigma}\pm k_{F}/K_{\rho}\right)$,
removes the coupling term in the Hamiltonian in Eq. (\ref{eq:H_eff_fsf})
allowing straightforward calculations of the observables.

The spectral function in Eq. (\ref{eq:A}) can be evaluated within
the framework of the effective field model in the same way. The original
Fermi operators $\psi_{\alpha}^{\dagger}\left(x\right)$ in the form
factor $\left|\left\langle f|\psi_{\alpha}^{\dagger}\left(0\right)|0\right\rangle \right|^{2}$
correspond to composite excitation consisting of two bosons (one for
spin and one for charge) and one fermion in the field language of
the model in Eq. (\ref{eq:H_eff_fsf}), see the state in Fig. \ref{fig:Spinfull-configuration}.
The fermionic part gives the dominant contribution to the spectral
weight, thus the spectral function reads $A\left(k,\varepsilon\right)=\int dtdxe^{i\left(kx-\varepsilon t\right)}\left\langle d^{\dagger}\left(x,t\right)d\left(0,0\right)\right\rangle $
where $d\left(x,t\right)=e^{-iHt}d\left(x\right)e^{iHt}$ and $\left\langle \dots\right\rangle $
is the zero temperature expectation value with respect to the model
in Eq. (\ref{eq:H_eff_fsf}). In the diagonal basis the average is
evaluated over free fields by standard means. Following the steps
of Ref. \onlinecite{Schmidt10} we obtain in proximity of the edge
$A\left(\varepsilon,k\right)\sim\theta\left(\varepsilon-\varepsilon_{\textrm{edge}}\left(k\right)\right)/\left|\varepsilon-\varepsilon_{\textrm{edge}}\left(k\right)\right|^{-\alpha}$
where the exponent depends only on a pair of the dimensionless Luttinger
parameters and the momentum along the spectral edge,\begin{widetext}
\begin{equation}
\alpha=\frac{1}{2}\pm\frac{1}{2}-\frac{K_{\rho}}{4}\left(1-\frac{\left(k-k_{F}\right)\left(\frac{k_{F}}{K_{\rho}^{2}}+\frac{k}{K_{\sigma}^{2}}\right)}{\left(\frac{k}{K_{\sigma}}\right)^{2}-\left(\frac{k_{F}}{K_{\rho}}\right)^{2}}\right)^{2}-\frac{K_{\rho}}{4}\left(\frac{1}{K_{\rho}}\pm\frac{\left(k-k_{F}\right)\left(\frac{k_{F}}{K_{\rho}K_{\sigma}}+\frac{k}{K_{\rho}K_{\sigma}}\right)}{\left(\frac{k}{K_{\sigma}}\right)^{2}-\left(\frac{k_{F}}{K_{\rho}}\right)^{2}}\right)^{2},\label{eq:alpha_fsf}
\end{equation}
\end{widetext} The result is different for the particle ($+$) and
the hole ($-$) sectors. The values of the Luttinger parameters obtained
numerically using Eq. (\ref{eq:v_fsf}), see Fig. \ref{fig:low_energy_fsf},
give divergent values of $0<\alpha<1$ in the particle sector and
cusp-like positive powers $-1<\alpha<0$ in the hole sector.

\section{Bosons}

While our primary interest lies in Fermi particles, for completeness
and to test the generality of our result we consider Bose particles
without spin. In this case the application of Bethe-Ansatz approach
is very similar to the case of Fermi particles without spin.\cite{Korepin}

We closely follow the approach of Lieb and Liniger in Ref. \onlinecite{LiebLiniger}.
In the coordinate basis a superposition of $N$ plain waves, $\Psi=\dotsint_{-\frac{L}{2}}^{\frac{L}{2}}dx_{1}\dots dx_{N}\allowbreak\sum_{P}e^{i\sum_{j}k_{P_{j}}x_{j}}\allowbreak e^{i\sum_{l<l'}\varphi_{P_{l}P_{l'}}}\allowbreak\psi^{\dagger}\left(x_{1}\right)\dots\psi^{\dagger}\left(x_{N}\right)\left|\textrm{vac}\right\rangle $,
is an eigenstate, $H\Psi=E\Psi$, of the model in Eq. (\ref{eq:H})
with the corresponding eigenenergy $E=\sum_{j=1}^{N}k_{j}^{2}/\left(2m\right)$.
Here the operators $\psi\left(x\right)$ obey the Bose commutation
rules $\left[\psi\left(x\right),\psi^{\dagger}\left(y\right)\right]=\delta\left(x-y\right)$,
$\sum_{P}$ is a sum over all permutation of $N$ quasimomenta $k_{j}$,
and the scattering phases $2\varphi_{ll'}=\log\Big[\left(k_{l}-k_{l'}+i2mU\right)\allowbreak/\left(k_{l}-k_{l'}-i2mU\right)\Big]/i$
are fixed by the two-body scattering problem. 

The periodic boundary condition quantises a set of $N$ quasimomenta
simultaneously 

\begin{equation}
k_{j}L-\sum_{l=1\neq j}^{N}2\varphi_{jl}=2\pi I_{j},\label{eq:Ij_bns}
\end{equation}
where $I_{j}$ are a set of non-equal integer numbers. The total momentum
of $N$ particles, $P=\sum_{j}k_{j}$, is a conserved quantity.

\begin{figure}[t]
\begin{center}\includegraphics[width=1\columnwidth]{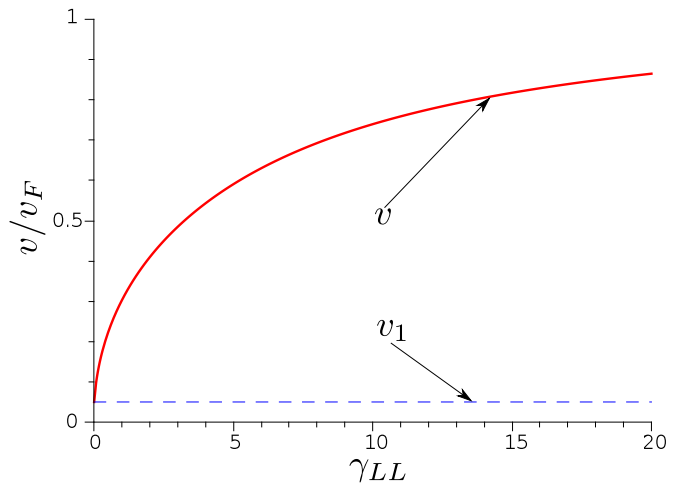}

\end{center}

\caption{\label{fig:low_energy_bns}The sound velocity of the collective modes
for spinless bosons $v$ at low energies as a function of the interaction
parameter $\gamma_{LL}$ from Eq. (\ref{eq:gammaLL}) - red line and
the quantum of momentum $v_{1}=2\pi/(mL)$ - blue dashed line; $L=200$,
$N=20$, $\gamma_{L}=0.2mU$.}
\end{figure}
The non-linear system of equations Eq. (\ref{eq:Ij_bns}) can be solved
explicitly in the limit of infinite repulsion. The hard-core bosons
in this limit are identical to free fermions\cite{TG} which decouples
Eq. (\ref{eq:Ij_bns}) into a set of plain wave quantisation conditions,
$k_{j}=2\pi I_{j}/L$. The corresponding eigenstates are Slater determinants
whose classification is identical to that of free fermions - all many-body
states correspond to all sets of $N$ non-equal integer numbers. These
values of quasimomenta $k_{j}$ can be adiabatically continued under
a smooth deformation of Eq. (\ref{eq:Ij_bns}) by varying the interaction
strength from $U=\infty$ to arbitrary value of $U$.

The single parameter that controls the behaviour of interacting bosons
can be obtained from the Bogoliubov theory in the weak interaction
regime.\cite{Bogoliubov47} This theory is valid when the interaction
length is smaller than the kinetic energy of particles, {\it e.g.} the high
density limit. The same parameter can be generalised to arbitrary
interaction strengths,\cite{LiebLiniger}

\begin{equation}
\gamma_{LL}=\frac{2mUL}{N}.\label{eq:gammaLL}
\end{equation}
When $\gamma_{LL}\ll1$ the interacting particles are like bosons
and when $\gamma_{LL}\gg1$ the system is almost a free Fermi (Tonks-Girardeau)
gas.

\begin{figure}[t]
\begin{center}

\includegraphics[width=1\columnwidth]{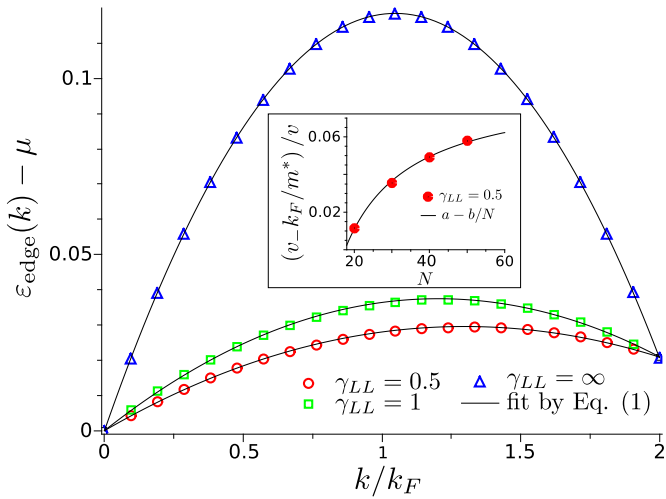}

\end{center}

\caption{\label{fig:Edge-for-spinless}Dispersion of the spectral edge mode
for spinless bosons for different values of the interaction parameter
$\gamma_{L}=0.5,1,\infty$; $L=200$, $N=20$. The blue triangles,
green squares, and red ellipses are the numerical solutions of Eq.
(\ref{eq:Ij_bns}), the solid black lines are the best parabolic fits
by Eq. (\ref{eq:edge_general}). Inset: Difference between the slope
of the parabolic dispersion at $E_{F}$, which is given by the effective
mass $m^{*}$, and the velocity of the sound modes, which is obtained
by direct evaluation of the energy of the first excited state above
the Fermi energy - $\left(v-k_{F}/m^{*}\right)/v$ - as a
function of the number of particles $N$ for $\gamma_{LL}=0.5$. The
solid black line is the $1/N$ fit, $a-b/N$, that gives $a=0.09$.}
\end{figure}
The eigenstates contributing to the spectral function in Eq. (\ref{eq:A})
satisfy the number of particles being constrained to be $N+1$. The
lowest energy state for a fixed momentum $-k_{F}+\Delta P$, where
$k_{F}=\pi N/L$, is given by the sets of integer numbers in Fig.
\ref{fig:spin-less-configuration}. At low energies the system is
well-approximated by the Tomonaga-Luttinger model with only two free
parameters\cite{Giamarchi}. Using the parameterisation in Fig. \ref{fig:spin-less-configuration},
the first Luttinger parameter (the sound velocity of the collective
modes) is a discrete derivative $v=L\left(E_{2}-E_{1}\right)/\left(2\pi\right)$,
where $E_{1}$ and $E_{2}$ are the energies of the states in Fig.
\ref{fig:spin-less-configuration} with $\Delta P=0$ and $\Delta P=2\pi/L$.
For Galilean invariant systems the second (dimensionless $K$) Luttinger
parameter can be obtained from the relation $vK=v_{F}$ where $v_{F}=\pi N/\left(mL\right)$.\cite{Haldane81}
Numerical evaluation of $v$ as a function of the interaction parameters
$\gamma_{LL}$ is given in Fig. \ref{fig:low_energy_bns}.

Beyond the linear regime the position of the edge of the spectral
function is given by the momentum dependence of the states in Fig.
\ref{fig:low_energy_bns}, $\varepsilon_{\textrm{edge}}\left(k\right)=E_{k}-E_{0}$
where $E_{k}$ corresponds to the states with $k=\Delta P$. Numerical
evaluation shows that the shape of $\varepsilon_{\textrm{edge}}\left(k\right)$
is close to a parabola for all values of $\gamma_{LL}$, see Fig \ref{fig:Edge-for-spinless}.
The biggest deviation from a parabola occurs when $\gamma_{LL}\ll1$.
We quantify it by comparing the effective mass $m^{*}$, obtained
by the best fit of Eq. (\ref{eq:edge_general}), with $v$ in Fig.
\ref{fig:low_energy_bns}, obtained at low energies. The deviation
$\left(v-k_{F}/m^{*}\right)/v$ increases as the number of particles
$N$ grows but it saturates at a finite value of $\sim0.1$ in the
limit $N\rightarrow\infty$, see inset in Fig. \ref{fig:Edge-for-spinless}. 

As with Fermi particles with spin, for finite systems the parabola-like
behaviour of the spectral edge mode breaks down in the ultra-weak
interaction regime when the sound velocity of collective modes at
low energies mode becomes comparable with its own quantum set by the
finite size of the system $v_1=2 \pi/\left(mL\right)$, see the dashed
line in Fig. \ref{fig:low_energy_bns}. Correspondingly, the threshold
for entering this regime becomes $\gamma_{LL}\rightarrow0$ in the
thermodynamic limit, as observed in Fig. \ref{fig:low_energy_bns}
when $v_{1}\rightarrow0$. When $v\sim v_1$ the edge
of the spectral function is linear at all energies, including the
high energy domain, with the slope that is governed by the kinetic
energy of a single free Bose particle.

\section{Conclusions}

In this work, we have analysed the spectral edge mode for a variety
of one-dimensional models with short-range interactions that bounds
from below a continuum of many-body excitations. Explicit diagonalisation
by means of Bethe-Ansatz techniques shows this mode to have an almost
perfect parabola dispersion in all cases. Based on this emergent phenomenon,
the spectral edge mode can be described empirically by a free, non-relativistic
particle with effective mass identified from the low energy theory
as a free electron mass strongly renormalised by interactions via
the dimensionless Luttinger parameter $K$ ($K_{\sigma}$ for particles
with spin). However, unlike a free particle, the spectral edge mode
is not protected by a symmetry thus deviations from the quadratic
dispersion may develop - the biggest discrepancy ($\lesssim20\%$)
occurs for Fermi particles with spin and a very large interaction
strength. The empirical model remains robust for finite sound velocities
of the collective modes at low energies $v\left(v_{\sigma}\right)>v_{1}$,
where $v_{1}=2\pi/(mL)$ is the quantum of momentum.

The relevance of the Luttinger (low energy) parameters beyond the
low energy limit implies that they can be extracted using a much wide
range of experimental data using the whole energy window from the
bottom of the band to the Fermi energy. However, the dispersion of
the spectral edge mode itself can not be used as a qualitative feature
to rule out interaction effects since the interactions between particles
do not change the parabolic shape of the single particle dispersion.
The biggest deviations could be observed for strongly interacting
spinful fermions ($K_{\sigma}\gtrsim10$), {\it e.g.} electrons in semiconductors
at low densities or cold Fermi atoms in a 1D trap that would require
a good resolution of the experiment.

The main result of this paper Eq. (\ref{eq:edge_general}) complements
the mobile impurity model which was developed by Glazman and co-workers
as a description of one-dimensional systems above the spectral edge
at high energies. Our explicit expression for the dispersion of the
edge mode removes an arbitrary input parameter (curvature of the dispersion)
that leaves only the few Luttinger parameters and the bare electron
mass as a minimal set of necessary ingredients to model excitations
above the spectral edge at arbitrary energies. Within such a framework,
for example, exponents of the spectral functions are expressed explicitly
in terms of only a few Luttinger parameters. The results in Eqs. (\ref{eq:alpha_fns},
\ref{eq:alpha_fsf}) provide a systematic way to classify the edge exponents
for wide range of microscopic parameters.
\begin{acknowledgments}
We thank EPSRC for the financial support through Grant No. EP/J016888/1.\end{acknowledgments}


\begin{thebibliography}{10}
\bibitem{Repulsion}In this paper we restrict ourselves to only repulse
interactions. An attractive potential admits formation of bound states
that can alter the properties of the system completely.
\bibitem{gamma_yg_gs}The definition of $\gamma_{YG}$ does not change
for excited states around the edge of the spectral function in the
thermodynamic limit. The average energy per magnon in Eq. (\ref{eq:gammaYG})
is still set by the ground state properties which are not altered
by a few excitations within the band of $\lesssim E_{F}$ in a large
system.
\bibitem{parabolic_fit_fsf}The biggest deviation of $20\%$ quantifies
the discrepancy between the best by Eq. (\ref{eq:edge_general}) at
high energy and the low energy velocity of the collective modes obtained
using Eq. (\ref{eq:v_fsf}). The fit by Eq. (\ref{eq:edge_general})
on its own is in fact more accurate: the coefficient of determination
is already $R^{2}=0.9934$ for $N=80$ in the strong interaction regime
with $\gamma_{YG}=6.88$.
\end{thebibliography}
\end{document}